\documentclass[conference]{IEEEtran}

\usepackage{balance}
\usepackage[hidelinks]{hyperref}

\IEEEoverridecommandlockouts
\usepackage{cite}
\usepackage{amsmath,amssymb,amsfonts}
\usepackage{algorithmic}
\usepackage{graphicx}
\usepackage{textcomp}
\usepackage{xcolor}
\def\BibTeX{{\rm B\kern-.05em{\sc i\kern-.025em b}\kern-.08em
    T\kern-.1667em\lower.7ex\hbox{E}\kern-.125emX}}
\begin{document}

\title{AGIR: Automating Cyber Threat Intelligence Reporting with Natural Language Generation}

\author{\IEEEauthorblockN{Filippo Perrina}
\IEEEauthorblockA{\textit{Department of Mathematics} \\
\textit{University of Padova}\\
Padua, Italy \\
filippo.perrina@studenti.unipd.it}
\and
\IEEEauthorblockN{Francesco Marchiori}
\IEEEauthorblockA{\textit{Department of Mathematics} \\
\textit{University of Padova}\\
Padua, Italy \\
francesco.marchiori@math.unipd.it}
\and
\IEEEauthorblockN{Mauro Conti}
\IEEEauthorblockA{\textit{Department of Mathematics} \\
\textit{University of Padova}\\
Padua, Italy \\
mauro.conti@unipd.it}
\and
\IEEEauthorblockN{Nino Vincenzo Verde}
\IEEEauthorblockA{\textit{Leonardo Cyber\&Security} \\
\textit{Leonardo S.p.A.}\\
Rome, Italy \\
nino.verde@leonardo.com}
}

\maketitle


\begin{abstract}
Cyber Threat Intelligence (CTI) reporting is pivotal in contemporary risk management strategies.
As the volume of CTI reports continues to surge, the demand for automated tools to streamline report generation becomes increasingly apparent. While Natural Language Processing techniques have shown potential in handling text data, they often struggle to address the complexity of diverse data sources and their intricate interrelationships.
Moreover, established paradigms like STIX have emerged as de facto standards within the CTI community, emphasizing the formal categorization of entities and relations to facilitate consistent data sharing.

In this paper, we introduce AGIR (Automatic Generation of Intelligence Reports), a transformative Natural Language Generation tool specifically designed to address the pressing challenges in the realm of CTI reporting.
AGIR's primary objective is to empower security analysts by automating the labor-intensive task of generating comprehensive intelligence reports from formal representations of entity graphs.
AGIR utilizes a two-stage pipeline by combining the advantages of template-based approaches and the capabilities of Large Language Models such as ChatGPT.
We evaluate AGIR's report generation capabilities both quantitatively and qualitatively.
The generated reports accurately convey information expressed through formal language, achieving a high recall value (0.99) without introducing hallucination.
Furthermore, we compare the fluency and utility of the reports with state-of-the-art approaches, showing how AGIR achieves higher scores in terms of Syntactic Log-Odds Ratio (SLOR) and through questionnaires.
By using our tool, we estimate that the report writing time is reduced by more than 40\%, therefore streamlining the CTI production of any organization and contributing to the automation of several CTI tasks.
\end{abstract}


\section{Introduction}
\label{sec:introduction}

The evolving cyber threat landscape has witnessed a dramatic surge in both the frequency and sophistication of attacks in recent years.
From traditional phishing emails to the stealthy and advanced operations of highly sophisticated cybercriminal groups known as Advanced Persistent Threats (APTs), the digital realm has become a battleground for organizations seeking to safeguard their data and infrastructure~\cite{chen2014study}.
To counter these evolving threats effectively, organizations have begun implementing the discipline of Cyber Threat Intelligence (CTI) to address them proactively.
CTI involves the systematic collection, analysis, and dissemination of data from diverse sources, including network logs, social media, and dark web forums, to identify, comprehend, and mitigate cyber threats.
By providing actionable insights into the Tactics, Techniques, and Procedures (TTPs) employed by cybercriminals, as well as the associated Indicators Of Compromise (IOCs), CTI can enhance an organization's situational awareness and reinforce its ability to detect and respond to attacks swiftly, ultimately reducing overall risk exposure~\cite{wagner2019cyber}.

An integral part of CTI is the production of comprehensive security reports.
Indeed, while several standards such as Structured Threat Information Expression (STIX) are employed to facilitate the sharing of structured data, natural language remains the most common and easily understandable format to collect and disseminate intelligence~\cite{barnum2012standardizing}.
These reports are a repository of detailed information about cyber threats, encompassing TTPs, exploited vulnerabilities, and IOCs.
Furthermore, they are vital in sharing CTI knowledge internally within organizations and externally with law enforcement agencies and other cybersecurity entities.
Unfortunately, the manual creation of these reports can be an exceptionally time-consuming and resource-intensive task.
Indeed, security analysts must aggregate and analyze extensive datasets before synthesizing their findings into clear and concise reports.
Additionally, the accurate reconstruction of a specific incident or a threat might need the collaboration of several analysts and the congregation of multiple intelligence sources, further aggravating the complexity of the task~\cite{oosthoek2021cyber}.
To tackle this challenge, Natural Language Generation (NLG) techniques have emerged as a promising solution for automating the report generation process~\cite{dale2020natural}.
Natural Language Generation models are already deployed in many instances to facilitate the production of textual data, such as financial summaries~\cite{plachouras2016interacting}, user tailoring and profiling in healthcare~\cite{balloccu2020nlg}, and chatbots~\cite{adamopoulou2020overview}.
Thus, NLG tools can save security analysts substantial time and resources by automating the conversion of structured data into well-crafted written content.
However, despite the evident advantages NLG offers, applying such tools within the realm of cybersecurity, specifically for generating security reports, remains partially unexplored.

\textbf{Contribution.}
This paper aims to address this gap in the cybersecurity literature by introducing \textbf{AGIR}, an NLG tool designed to automate the creation of cybersecurity reports from structured data.
AGIR leverages STIX graphs, constituted by threat entities and their relationships, to parse them with a template-based approach and subsequently leverage a Large Language Model (LLM) to improve the fluency and utility of the generated report.
While AGIR currently supports four distinct report types, its pipeline is designed with extensibility in mind, allowing for the incorporation of additional report formats as needed.
The quantitative evaluation of our approach shows that the information in the structured data is conveyed in the generated reports with almost perfect recall values (0.99) without introducing any hallucination in the process.
Furthermore, we qualitatively evaluate the outputs of our tool through the Syntactic Log-Odds Ratio (SLOR) metric and questionnaires answered by experienced cyber threat analysts.
The results show that AGIR produces fluent reports that outclass state-of-the-art models while also maintaining a high level of utility and reducing report writing time by 42.6\%.

Our contributions can be summarized as follows.
\begin{itemize}
    \item We introduce \textbf{AGIR}, a Natural Language Generation (NLG) tool designed to automate the creation of cybersecurity reports from structured data.
    \item We propose a pipeline design aimed at enhancing the scalability of our tool and enabling the inclusion of a wider range of supported report types.
    \item We assess AGIR's performance through a combined quantitative and qualitative approach, utilizing several metrics and conducting surveys with expert threat analysts.
    \item We make several samples of the generated reports available at \url{https://github.com/Mhackiori/AGIR}.
\end{itemize}

\textbf{Organization.}
The paper is organized as follows.
In Section~\ref{sec:background}, we introduce key CTI and NLG concepts, setting the stage for our discussion.
In Section~\ref{sec:relatedworks}, we explore the use of Natural Language Processing (NLP) and NLG techniques in the cybersecurity domain.
Section~\ref{sec:systemmodel} gives an overview of the system model and the possible implementations of AGIR.
Section~\ref{sec:methodology} delves into the technical specification of AGIR, offering insights into its pipeline.
In Section~\ref{sec:evaluation}, we evaluate AGIR's performance both quantitatively and qualitatively through human evaluation and SLOR metric.
Finally, Section~\ref{sec:conclusions} concludes this work.
\section{Background}
\label{sec:background}

In this section, we give a more thorough background on the techniques and notions that we use in the methodology.
In particular, we focus on Cyber Threat Intelligence and the type of data that this discipline deals with (Section~\ref{subsec:cti}) and the core concepts of Natural Language Generation (Section~\ref{subsec:nlg}).

\subsection{Cyber Threat Intelligence}
\label{subsec:cti}

CrowdStrike, an American security company, defines Threat Intelligence as the process of collecting, processing and analyzing data to comprehend threat actors' motives, targets, and attack behaviors.\footnote{\url{https://www.crowdstrike.com/cybersecurity-101/threat-intelligence/}}
This intelligence empowers quicker, data-driven security decisions, shifting from a reactive to a proactive stance against threat actors.
Threat intelligence sources encompass open source data, social media, device logs, internet traffic, and information from the deep and dark web.
In today's cybersecurity landscape, it is pivotal in enabling organizations to proactively identify and mitigate potential threats, making them more resilient against cyberattacks.

Threat intelligence can be divided into three main categories: strategic intelligence (intended for non-technical audiences and offering high-level insights into threats and vulnerabilities), tactical intelligence (intended for technically prolific teams and providing immediate threat indicators), and operational intelligence (intended for professionals and offering in-depth insight into TTPs)~\cite{keim2019cyber}.
Additionally, each piece of intelligence undergoes a defined life cycle, spanning from planning and direction to dissemination and integration~\cite{porkorny2018phases}.

\begin{figure}[!htpb]
    \centering
    \includegraphics[width=\linewidth]{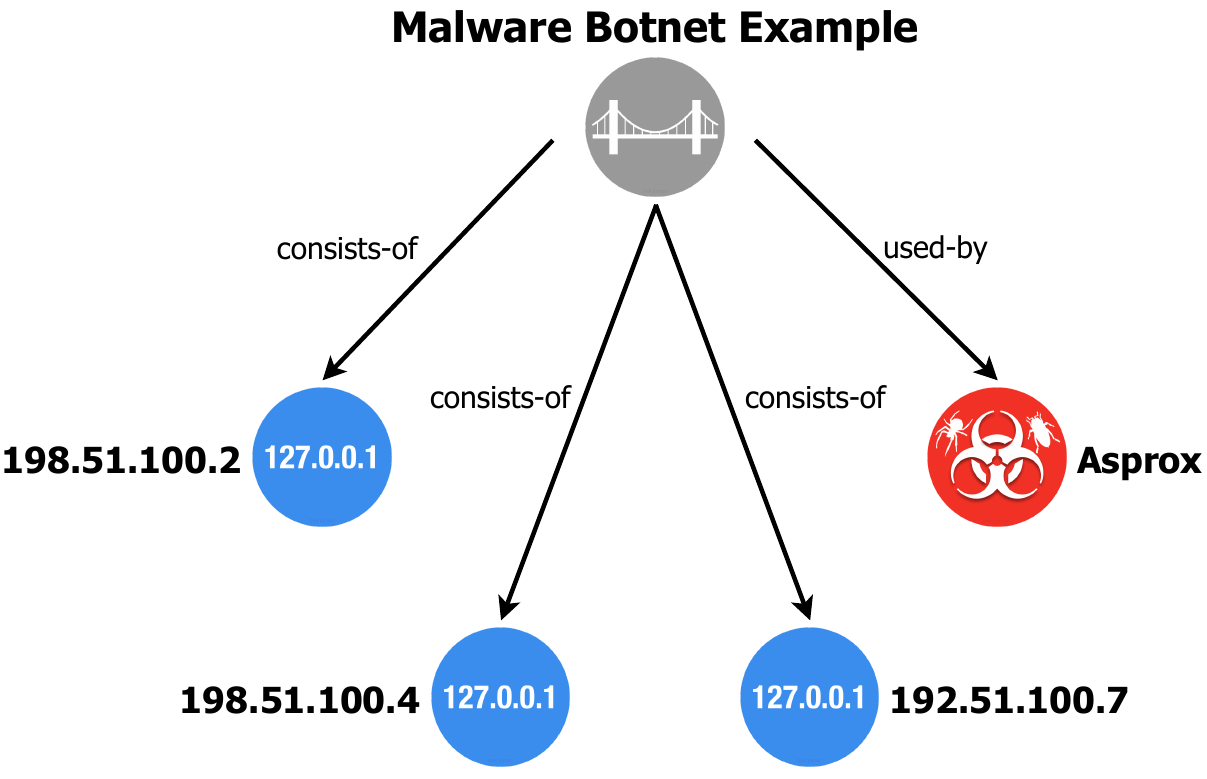}
    \caption{Graphical example of a STIX graph. Icons used in this paper are from the 'stix2-graphics' repository by Bret Jordan~\cite{githubGitHubFreetaxiistix2graphics}.}
    \label{fig:stix}
\end{figure}

Each of the intelligence types can be shared among organizations in various formats.
While natural language reports are the most common medium when dealing with elaborate threats, structured data standards have been created to facilitate machine-readability and ease of dissemination.
The most common standard used in CTI is STIX (Structured Threat Information Expression), which includes several types of entities and relationships that allow for a graph representation of the intelligence~\cite{barnum2012standardizing}.
Effectively, with STIX, intelligence is shared in the form of JSON files that can be represented as a connected graph of nodes and edges, in which each node represents an entity and each edge represents a relationship between entities.
A graphical overview of one of these graphs is shown in Figure~\ref{fig:stix}.
These JSON files constitute part of AGIR's input.

\subsection{Natural Language Generation}
\label{subsec:nlg}

Natural Language Generation (NLG) is a subset of Natural Language Processing (NLP) centered on creating computer systems capable of generating human-like language output, such as documents and reports~\cite{dale2020natural}.
NLG can be categorized into text-to-text and data-to-text applications.
Text-to-text processes utilize existing texts to generate coherent new text, while data-to-text systems transform non-linguistic structured data, like tables or graphs, into natural language text.
Our focus is on data-to-text generation, as it aligns with the contribution of this paper.

Data-to-text generation approaches fall into three categories.
\begin{itemize}
    \item \textbf{Rule-based} -- These methods follow a three-stage pipeline, starting with content selection and text structuring in the Document Planner~\cite{mcdonald1993issues}.
    The Microplanner combines sentence aggregation, lexicalization, and referring expression generation~\cite{reiter1997building}.
    Finally, the Linguistic Realizer generates grammatically correct sentences~\cite{gatt2018survey}.
    \item \textbf{Template-based} -- Unlike rule-based approaches, template-based methods directly map non-linguistic input to linguistic surface structure.
    Text generation occurs through string manipulation, where users create programs with string patterns containing empty slots to be filled with relevant information.
    \item \textbf{Neural-based} -- Neural approaches are data-driven and require no manual feature engineering.
    They develop end-to-end models where neural networks learn to generate high-quality text descriptions directly from input data, bypassing explicit modeling of intermediate stages.
\end{itemize}

In the early stages of NLG, rule-based systems were favored for embodying linguistic insights~\cite{reiter1995nlg}.
This perception shifted with the development of more sophisticated template-based systems like D2D, which could adapt output based on context and perform complex syntactic operations~\cite{theune2001data}.
Template-based systems offer ease of development, control, and speed but may sacrifice fluency, maintainability, and flexibility compared to rule-based systems.
In modern NLG, neural-based approaches have taken precedence, thanks to their superior generalization and output variation capabilities.
However, neural-based systems might lack control over the generation process, potentially resulting in inaccurate text generation.
They also require extensive training data, limiting their feasibility when large datasets are unavailable.
Hybrid approaches, which combine two or more methods, aim to leverage the strengths of each approach.
For instance, Kale and Ratsogi implemented a two-stage pipeline combining a template-based approach for a baseline response and a pre-trained language model (T5) for rewriting the response into coherent, natural-sounding text~\cite{kale2020template}.
This hybrid method blends the control and ease of development from templates with the fluency and diversity offered by neural approaches.
\section{Related Works}
\label{sec:relatedworks}

In this section, we overview related works on applying Natural Language Processing and Natural Language Generation in the cybersecurity domain, respectively, in Section~\ref{subsec:nlpcti} and Section~\ref{subsec:nlgcti}.

\subsection{NLP for CTI}
\label{subsec:nlpcti}

While the use of Natural Language Generation in Cyber Threat Intelligence applications has not been thoroughly explored in the literature, Natural Language Processing has been extensively researched for Information Extraction (IE) purposes.
Indeed, automatically processing CTI reports to retrieve entities and relationships can be pivotal for an organization seeking to gather intelligence with minimal time delays and manpower.
Effectively, these systems drive researchers towards fully automating CTI reporting, making cyber defense practices more accessible to all organizations.
One of these systems is STIXnet, which deploys several IE models to extract all STIX entities and relationships from natural language reports~\cite{10.1145/3600160.3600182}.
By combining IE systems such as STIXnet with AGIR, it will be possible to automatically process a multitude of reports and parse their intelligence in one single report, tailored according to the user's needs.

\subsection{NLG for CTI}
\label{subsec:nlgcti}

To the best of our knowledge, there is only one NLG approach applied to CTI in a setting similar to ours, while only a few instances are applied to the cybersecurity domain.
One of these few examples involves Das and Varma's work, who developed a system employing Recurrent Neural Networks (RNN) to generate text for advanced email masquerading~\cite{das2019automated}.
Other instances instead include implementing NLG models for cybersecurity education and training purposes~\cite{skrodelis2023latest}.
The usage of NLG models for CTI report generation has also been explored by Ranade et al. for poisoning attack purposes~\cite{ranade2021generating}.
In their paper, the authors use GPT-2 to generate fake CTI text to poison the dataset of Cybersecurity Knowledge Graphs.
However, while this paper also treats the problem from the perspective of an attacker, it also utilizes textual inputs and thus differs from our application, which uses structured data as input.
The research work that deals with a setting similar to ours is the one by S. Polzunov and J. Abraham, where they introduced Narrator, a tool capable of creating intelligence reports from the JSON representation of STIX graphs~\cite{abraham2020narrator}.
Narrator employs a rule-based approach to generate four types of reports, which can be edited and exported in PDF format.
AGIR expands upon the foundation laid by Narrator, enhancing overall performance.
In its first pipeline step, AGIR adopts a template-based approach inspired by the D2D  system.
The second step incorporates a technique akin to that employed by Kale and Ratsogi, utilizing ChatGPT to rewrite reports in a more human-like manner.
\section{System Model}
\label{sec:systemmodel}

As an automatic CTI report generation system, AGIR leverages the STIX graphs of entities and relationships to fulfill its objective.
Thus, to maximize its capabilities, an underlying Knowledge Base (KB) can be deployed to store the STIX data.
The usage of a KB yields several advantages.

\begin{itemize}
    \item \textbf{Intelligence Categorization} -- Entities and relations can be stored and categorized using the STIX guidelines.
    \item \textbf{Incident Recostruction} -- Several sources of intelligence can be parsed together to generate a single report containing the most amount of information on a specific threat or incident.
    \item \textbf{Threat Evolution} -- Intelligence collected over a period of time can be aggregated to construct a timeline, providing insights into the evolution of a specific threat.
\end{itemize}

Several sources can be employed to collect the intelligence stored in the KB.
One example is the MITRE ATT\&CK framework, which publicly stores and updates intelligence on threat groups, TTPs, mitigations, and software~\cite{strom2018mitre}.
Through the usage of APIs, it is also possible to collect the intelligence automatically in a structured format, which can then be converted to STIX.
Reports and CTI bulletins can also be used as a starting point for populating the KB.
Indeed, as anticipated in Section~\ref{subsec:nlpcti}, IE systems can be deployed for this purpose, making CTI reporting almost fully automatic.

\begin{figure}[!htpb]
    \centering
    \includegraphics[width=\linewidth]{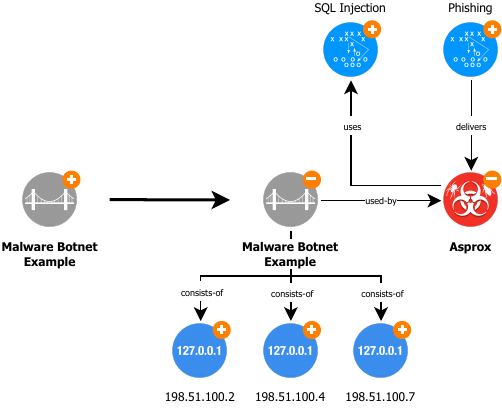}
    \caption{Proof-of-concept of AGIR's GUI.}
    \label{fig:gui}
\end{figure}

The storage of entities, relations, and the reports from which the intelligence has been collected allows for a more flexible implementation of AGIR.
Suppose the generation of a report on a specific infrastructure is requested.
The system can automatically query the Knowledge Base with all the intelligence related to that specific entity, reporting attributes, dates, and other information that can be of use.
Users can dynamically select the amount of intelligence they want to include by using the Graphical User Interface (GUI) of the service, expanding nodes on the graphs as shown in Figure~\ref{fig:gui}.

\begin{figure}[!htpb]
    \centering
    \includegraphics[width=.65\linewidth]{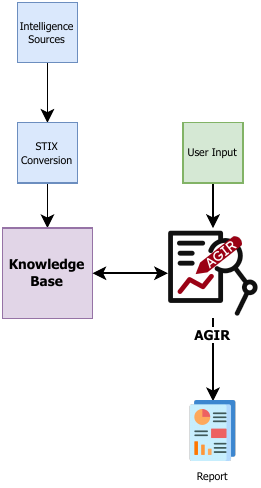}
    \caption{Overview of the system model.}
    \label{fig:system}
\end{figure}

In this example, we use the same dummy data used for Figure~\ref{fig:stix}.
We first consider the ``Malware Botnet Example'' infrastructure as a singular entity, which will constitute the main subject of the report.
By expanding the node, all the information about IPv4 addresses and malware previously shown in Figure~\ref{fig:stix} appear.
We can further expand those nodes, as it is possible to see for the ``Asprox'' entity, which will introduce two attack patterns in the graph.
The selected STIX graph can then be fed as input to AGIR, creating a CTI report accordingly.
Figure~\ref{fig:system} shows an overview of the overall system model.

\section{Methodology}
\label{sec:methodology}

\begin{figure*}[!htpb]
    \centering
    \includegraphics[width=\linewidth]{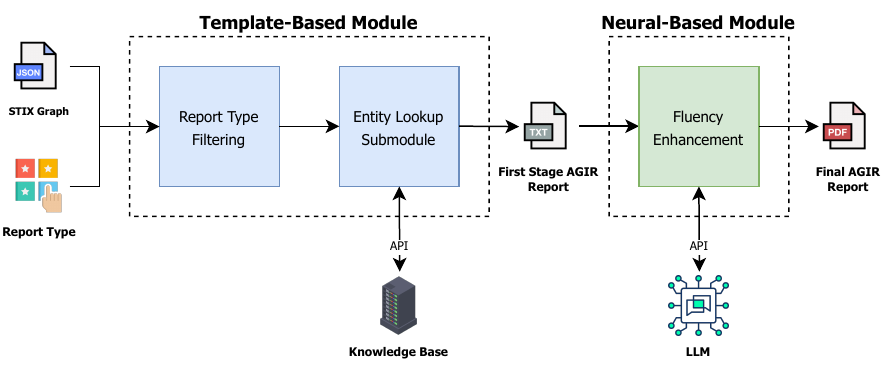}
    \caption{Overview of AGIR's pipeline.}
    \label{fig:pipeline}
\end{figure*}

In this section, we overview the methodology that constitutes AGIR and the modules that allow for generating CTI reports.
The template-base module and the neural-based are presented, respectively, in Section~\ref{subsec:template} and Section~\ref{subsec:neural}.
An overview of AGIR's pipeline is shown in Figure~\ref{fig:pipeline}.

\subsection{Template Based Module}
\label{subsec:template}

The template-based module is AGIR's first module and, thus, inherits its input, which is constituted as follows.
\begin{enumerate}
    \item \textbf{STIX Graph} -- A JSON file representing the graph from which to generate the report.
    \item \textbf{Report Type} -- A parameter given by the user indicating the type of report that the system should generate.
\end{enumerate}

In the initial stage of the template-based module execution, we analyze the report type parameter and designate the template accordingly.
Then, depending on the selected template, we select a subset of entities and relationships from the JSON representation of the previously mentioned graph.
This is done because of the specific type of intelligence that each report template contains.
Thus, by first considering the template selection, we reduce the number of API calls that we need to perform to retrieve the entities from the Knowledge Base.
Indeed, our system supports four different report types.
\begin{enumerate}
    \item \textbf{Overview Report} -- Provides a summary of information from the STIX graph.
    Analysts can use this template to obtain a clear understanding of a specific incident portrayed in the JSON file.
    \item \textbf{Subject Report} -- This template focuses on specific entities such as threat actors or intrusion sets.
    It also includes relationships, IOCs, and MITRE sections for all other selected entities.
    This type of intelligence helps understand the landscape where the subject is inserted.
    \item \textbf{Timeline Report} -- Provides a timeline overview of the entities related to the STIX graph.
    This template sorts the events chronologically and reports them according to their sequencing.
    \item \textbf{Vulnerability Report} -- Deals with vulnerabilities related to a specific entity and thus constitutes the most specific template out of the four.
    For each vulnerability, it also includes a table showing specific properties such as CVSS (Common Vulnerability Scoring System) score, mitigations, and vulnerable configurations. 
\end{enumerate}
After the template selection, for each entity in the STIX graph, we extract its type from the Knowledge Base, which will be retrieved in conjunction with its unique identifier.
Using IDs allows for efficiently storing reports, relationships, and events associated with a specific entity.
Thus, each time an entity is called from the Knowledge Base, we can also retrieve all its history and the previous relationships it had with other entities.
To do so, for each selected entity from the STIX graph, we initialize a dictionary with the following six keys: \textit{overview}, \textit{relationships}, \textit{stats}, \textit{useful resources}, \textit{IOCs}, and \textit{TTPs}.
Based on the entity ID, we then query the KB through several API calls to populate the dictionary accordingly.
This means performing content selection based on a predefined set of rules (e.g., properties specific to the entity are inserted in the overview section, and information about related IOC is inserted in the IOC section).
Once all the information is available, the module goes through each section of the report template and fills the gaps with the appropriate piece of intelligence.


\subsection{Neural Based Module}
\label{subsec:neural}

While the output of the template-based module can already be considered as a report, being generated from a set of rules implies that it is naturally mechanic and not fluent.
Thus, to improve on this aspect, we use the neural-based module.
To do so, we use ChatGPT, a Large Language Model that, in recent months, has attracted an incredible amount of attention due to its efficiency in Natural Language Generation~\cite{mijwil2023towards}.
Through its APIs, we prompt the model for a more fluent version of the report, highlighting the need for keeping the information in the text unchanged.
Once the generation is complete, the final result is given as the output of the overall AGIR pipeline.

Two main challenges arise from using ChatGPT as a "fine-tuning" model for the report.
\begin{itemize}
    \item \textbf{Cost} -- While using the LLM through its graphical interface is free for all registered users, the payment of a fee is needed for each API call.
    At the time of the development of AGIR, the cost of generating a single report is, on average, 0.0024 US dollars.
    Since, in real-world scenarios, the number of reports generated each day is not exceptionally high, the price does not constitute an obstacle for companies and organizations.
    \item \textbf{Lack of Control} -- As stated in Section~\ref{subsec:nlg}, neural-based NLG approaches present a lack of control over the generated output.
    To still assess the contribution of this module on the overall fluency of the report, in Section~\ref{subsec:qualitative}, we qualitatively evaluate the outputs of both modules, proving the benefits of ChatGPT in the report generation.
\end{itemize}
\section{Evaluation}
\label{sec:evaluation}

This section presents an experimental evaluation of our system.
Our aim is to write a precise human-like report that supports analysts and reduces the time they spend in the report-writing process.
Given the importance of both content and style in the generation of a report, we split the evaluation into two parts.
In the first, we evaluate AGIR quantitatively, thus assessing the completeness of intelligence contained in the output (Section~\ref{subsec:quantitative}).
In the second, we evaluate the style of those reports in terms of fluency and utility (Section~\ref{subsec:qualitative}).

\subsection{Quantitative Results}
\label{subsec:quantitative}

We assess AGIR's accuracy to determine whether the incorporation of ChatGPT has led to any instances of omission or hallucination in the generated text.
It's worth noting that in this analysis, we do not evaluate the initial step of AGIR singularly (i.e., reports generated by the template-based module).
This is because the module relies entirely on predefined rules, and unless there are implementation errors, they are not expected to introduce accuracy concerns.

For the quantitative evaluation, we will use True Positives (TP), False Positives (FP), and False Negatives (FN), which in this specific application are defined as follows.
\begin{itemize}
    \item \textbf{True Positive} -- Information that is present both in the final report and in the input JSON file.
    \item \textbf{False Positive} -- Information that is present in the report but not in the input JSON file.
    \item \textbf{False Negative} -- Information that is present in the input JSON file but is not found in the report.
\end{itemize}
By using these indicators, we use three different metrics to evaluate AGIR accuracy: precision, recall, and F1-score.
These metrics are defined as follows.
\begin{equation}
\label{eq:precision}
    Precision = \frac{TP}{TP + FP},
\end{equation}
\begin{equation}
\label{eq:recall}
    Recall = \frac{TP}{TP + FN},
\end{equation}
\begin{equation}
\label{eq:f1}
    F1 = 2\frac{Precision \cdot Recall}{Precision + Recall}.
\end{equation}

To evaluate our system, we use a sample of 12 STIX graphs.
From each of these graphs, we execute AGIR and generate 12 reports, split equally between each report type.
Each JSON file underwent a manual process in which the intended report information was identified and subsequently verified for its presence within the generated report.
Results of this evaluation are shown in Table~\ref{tab:quantitative}.
As evident from the results, the utilization of ChatGPT consistently avoided introducing new information (i.e., hallucination events) into the report.
However, in a very limited number of cases, the model did omit information from the JSON file.
Nonetheless, having a recall and F1 score value close to 1 means that the generated reports almost always contain the same amount of information from the JSON file, highlighting the completeness of the reports.

\begin{table}[!htpb]
    \caption{AGIR's quantitative evaluation results.}
    \begin{center}
        \begin{tabular}{|c|c|c|}
            \hline
            \textbf{Precision}&\textbf{Recall}&\textbf{F1 Score}\\ \hline
            1.000&0.993&0.996\\ \hline
        \end{tabular}
    \label{tab:quantitative}
    \end{center}
\end{table}

\subsection{Qualitative Results}
\label{subsec:qualitative}

We now perform a qualitative evaluation of the reports generated by AGIR.
Thus, we now focus on fluency, correctness, and utility of the information expressed in natural language.
To do so, we further divide the evaluation into two parts: a syntactic evaluation (Section~\ref{subsub:syntactic}) and a linguistic evaluation (Section~\ref{subsub:linguistic}).

\subsubsection{Syntactic Evaluation}
\label{subsub:syntactic}

For the syntactic evaluation of the generated reports' qualitative properties, we introduce SLOR (Syntactic Log-Odds Ratio)~\cite{kann2018sentence}.
We opted for this metric because it is widely recognized as the de facto standard for assessing text fluency in referenceless evaluations, which aligns with our specific evaluation context.
SLOR exhibits the strongest correlation with human sentence acceptability compared to various sentence probability-based scoring methods~\cite{lau2017grammaticality}.
Furthermore, its effectiveness has been demonstrated in unsupervised text compression tasks.
SLOR assigns a score to each sentence $S$ by calculating its log probability using a specific Language Model (LM).
This score is then normalized by the log probability of unigrams and the sentence's length.

\begin{equation}
\label{eq:slor}
    SLOR \left( S \right) = \frac{1}{S} \left( \ln \left( p_M \left( S \right) \right) \right) - \ln \left( p_u \left( S \right) \right).
\end{equation}
In Equation~\ref{eq:slor}, $p_M\left(S\right)$ is the probability assigned to the sentence under the LM, while $p_u\left(S\right)$ is the unigram probability of the sentence $S$.
These quantities are defined, respectively, as follows.

\begin{equation}
\label{eq:pm}
    p_M \left( S \right) = p \left( \langle t_1, t_2, ..., t_{\left| S \right|} \rangle \right) = p\left( t_1 \right) \prod^{\left| S \right|}_{i = 2} p \left( t_i | t_1, ..., t_{i-1} \right),
\end{equation}
\begin{equation}
\label{eq:pu}
    p_u \left( S \right) = \prod_{t \in S} p \left( t \right).
\end{equation}
The rationale behind subtracting unigram log probabilities is to mitigate the impact of a token's rarity when considered individually, as opposed to its rarity in a specific sentence position.
Normalizing by sentence length is essential to ensure that shorter sentences are not favored over equally fluent longer ones.
It is worth noting that the log probability of a sentence normalized by its length corresponds to the negative cross-entropy of that sentence, as per the employed language model during the evaluation.
To calculate sentence probabilities, we utilize the pre-trained XLNet language model~\cite{yang2019xlnet}.
SLOR scores are theoretically unbounded, with higher scores indicating better text fluency.
The value range depends on many factors, such as the dataset, language model, and the nature of the text being evaluated.
Thus, we are interested in comparing the results from our testbed with other state-of-the-art NLG models applied on CTI, such as Narrator.
Furthermore, we also evaluate the fluency of the reports in different stages of the pipeline, thus assessing the contribution of each module.
Therefore, the three models that we evaluate are the following: Narrator, first step AGIR (i.e., the output of the template-based module), and final AGIR (i.e., the output of the neural-based module).
The evaluation dataset is the same as the one used for the quantitative evaluation in Section~\ref{subsec:quantitative}, and thus comprises 3 STIX graphs for each report type supported by AGIR.
Results are shown in Table~\ref{tab:slor}.
As we can see, the SLOR score obtained by Narrator and first stage AGIR are very close to one another, while final AGIR has a significant increase in the score.
These results confirm not only the contribution that the neural-based module has on the overall quality of the reports but also highlight the possible limitations of template-based approaches when considered independently.

\begin{table}[!htpb]
    \caption{Average SLOR scores and standard deviation of Narrator and different stages of AGIR's pipeline.}
    \begin{center}
        \begin{tabular}{|c|c|}
            \hline
            \textbf{Model}&\textbf{SLOR}\\ \hline
            Narrator&2.13$\pm$0.90\\ \hline
            First Step AGIR&2.16$\pm$1.07\\ \hline
            \textbf{Final AGIR}&\textbf{2.75$\pm$0.72}\\ \hline
        \end{tabular}
    \label{tab:slor}
    \end{center}
\end{table}

\subsubsection{Linguistic Evaluation}
\label{subsub:linguistic}

By using SLOR, we can assess the fluency of each report with formally defined metrics, thus giving an objective perspective on AGIR's efficacy.
However, we are also interested in the human perspective, given our system's most important feature: the reduction of processing times for CTI reporting.
For this reason, we conduct another type of qualitative evaluation based on the opinions shared by expert threat analysts.
All questioned analysts are employers of Leonardo S.p.A., an Italian multinational company that collaborated in this research.
Opinions were collected in the form of surveys, which involved rating the quality of the reports based on the following three dimensions.
\begin{itemize}
    \item \textbf{Fluency} -- Whether the text is easy to read and understand.
    \item \textbf{Correctness} -- Whether the content of the text is true and derivable from the input data.
    \item \textbf{Utility} -- Whether the text helps the user to write the final report faster.
\end{itemize}
The questionnaire collected opinions on all of these aspects from 38 analysts.
Fluency and correctness have been measured using a Likert scale from 1 (not good) to 5 (very good)~\cite{likert1932technique}.
The utility dimension instead is evaluated by asking each analyst how long they think it will take for them to write a final report starting from the output of the system.
The utility value is then compared with a baseline score of 127.3 minutes, which was obtained by asking all analysts how much time, on average, it takes for them to write a report.
To prevent the outcomes from being biased by a single questionnaire instance, we have devised three questionnaires with identical structures and distributed the analysts evenly into three groups.
Furthermore, the presented reports are unmarked, and therefore, analysts cannot know from which model they are generated.
The questionnaires consist of four sections, each corresponding to a report type and containing one report from each system, resulting in a total of 12 reports for each questionnaire.
Each report is assessed based on the three dimensions mentioned earlier.
In Table~\ref{tab:survey}, we show the obtained results from the survey.
As we can see, we also notice that first step AGIR still outclasss Narrator on every metric.
Moreover, final AGIR reports outperform the other models used for comparison in all three dimensions.
Fluency is the metric that is most impacted by the neural-based module, confirming our hypothesis and thus further asserting its contribution to the report generation process.
An increase in correctness from first step AGIR to final AGIR stands out from this evaluation, dispelling previous uncertainty about possible omissions or hallucinations that could have appeared with the usage of LLMs.
As for the model's utility, analysts believe that AGIR can reduce report production times by 42.6\% with respect to the baseline, resulting in a total time reduction of 54 minutes.
It should be noted that the results shown here are averaged on all the report types.
A more detailed analysis can be found in Appendix~\ref{app:questionnaires}.

\begin{table}[!htpb]
    \caption{Questionnaire results grouped by dimension.}
    \begin{center}
        \begin{tabular}{|c|c|c|c|}
            \hline
            \textbf{Model}&\textbf{Fluency}&\textbf{Correctness}&\textbf{Utility}\\ \hline
            Narrator&2.98&3.00&97.5 min\\ \hline
            First Step AGIR&3.48&3.77&79.6 min\\ \hline
            \textbf{Final AGIR}&\textbf{4.13}&\textbf{3.90}&\textbf{74.3 min}\\ \hline
        \end{tabular}
    \label{tab:survey}
    \end{center}
\end{table}
\section{Conclusions}
\label{sec:conclusions}

Cyber Threat Intelligence is an important topic for all companies and organizations.
With the use of CTI, defenses against threat actors can be built proactively and more efficiently, thus increasing the security of each asset.
However, the sharing of CTI data is still anchored to natural language, which further delays the integration of the intelligence and the application of the defense mechanisms.
Furthermore, the dissemination of the information is extremely time-consuming, forcing analysts to write several reports each day to distribute their data.

\textbf{Contribution.}
In this paper, we presented \textbf{AGIR}, a system for the Automatic Generation of Intelligence Reports.
AGIR takes in input the JSON representation of a STIX graph and uses it to generate a report containing all entities and relationships contained.
AGIR uses two different approaches for Natural Language Generation: a template-based module used for building a baseline report and a neural-based module used to increase its fluency.
Using this pipeline, we allow for the generation of four different types of reports (overview, subject, timeline, and vulnerability).
We experimentally evaluated AGIR both quantitatively and qualitatively.
We showed that reports almost perfectly include the entities and relationships contained in the STIX graph given in input (recall value of 0.99) without introducing any hallucination (precision value of 1).
Also, we assessed through Syntactic Log-Odds Ratio scores and questionnaires that our reports are more fluent with respect to other state-of-the-art models and that the usage of AGIR can reduce production times by 42.6\%.

\textbf{Future Works.}
For future research endeavors, we aim to validate the accuracy of the extrinsic evaluation results by involving more domain experts who can assess AGIR in the context of crafting actual security reports.
Presently, AGIR relies on ChatGPT, which may raise concerns related to cost and privacy.
An intriguing avenue for further investigation would be to explore the development of an equivalent deep learning model that can be employed locally, thereby mitigating the aforementioned issues.
Additionally, a valuable contribution could involve the creation of an extensive dataset containing the pertinent STIX properties for a range of entities.
Such a dataset could serve as the foundation for training a language model to generate initial reports, thus addressing the maintainability challenge associated with template usage.

\balance
\bibliographystyle{ieeetr}
\bibliography{references}

\clearpage
\nobalance
\appendices

\section{Questionnaires}
\label{app:questionnaires}

In this appendix section, we give more details on the questionnaires given to the expert analysts for the qualitative evaluation of AGIR.
In Appendix~\ref{app:samples}, we show a sample of the questionnaires and the reports provided to the analysts.
Moreover, we divide the evaluation of each model on each report type and show their results in Appendix~\ref{app:reporttype}.

\subsection{Samples}
\label{app:samples}

The survey was conducted through a Google Forms module that was sent to Leonardo's employees.
In particular, we focused on the staff members of the Security Operation Center (SOC), given their experience in CTI and, in particular, in threat intelligence reporting.
The questionnaires given to the analysts are structured as follows.
First, to establish a baseline to evaluate AGIR's utility, we ask each expert how much time, on average, it takes for them to write a full report.
Afterward, for each of the four supported report types, we do the following.

\begin{figure}[!htpb]
    \centering
    \includegraphics[width=\linewidth]{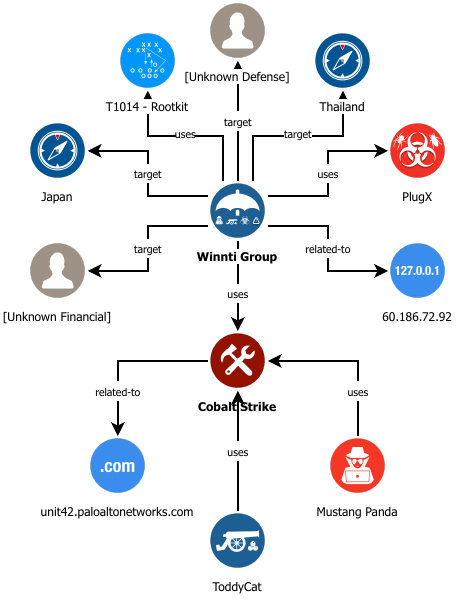}
    \caption{JSON input example for Overview and Subject reports. When dealing with the Subject report, we focus on the ``Winnti Group'' entity.}
    \label{fig:overview-subject}
\end{figure}

\begin{figure}[!htpb]
    \centering
    \includegraphics[width=\linewidth]{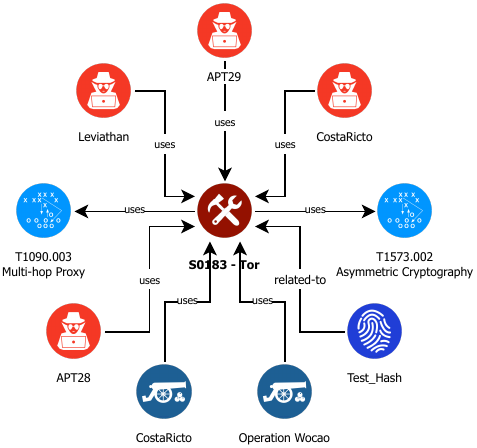}
    \caption{JSON input example for Timeline reports.}
    \label{fig:timeline}
\end{figure}

\begin{figure}[!htpb]
    \centering
    \includegraphics[width=\linewidth]{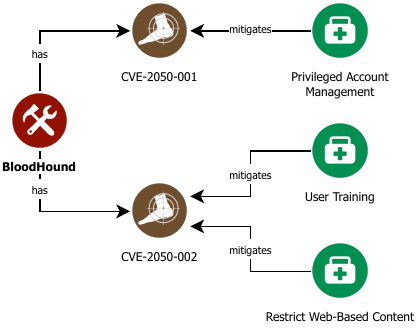}
    \caption{JSON input example for Vulnerability reports.}
    \label{fig:vulnerability}
\end{figure}

\begin{enumerate}
    \item We describe the aim of the report type and the focus of their entities and relationships.
    We also provide several use cases to further contextualize the usage of those particular reports.
    \item We provide a graphical overview of the JSON input from which the reports have been generated.
    Some examples of those STIX graphs are shown in Figure~\ref{fig:overview-subject}, Figure~\ref{fig:timeline}, and Figure~\ref{fig:vulnerability}.
    \item We provide three examples of generated reports.
    Those three samples are the report generated by Narrator, the report generated by first step AGIR, and the report generated by final step AGIR.
    To avoid biases, the samples are unmarked and randomized so that the analysts do not know which report has been generated by our system.
    \item We ask for an evaluation of each report's accuracy, described as the presence of true intelligence derivable from the input.
    \item We ask for an evaluation of each report's fluency, described as the quality of the text, its clarity, ease of reading, and how close it is to a human-generated report.
    \item We ask for an evaluation of each report's utility, described as how long the analyst would take to write a full report starting from the presented one.
\end{enumerate}
We provide several samples of the generated reports in our publicly available repository.\footnote{\url{https://github.com/Mhackiori/AGIR/tree/main/Reports}}

\subsection{Results Divided by Report Type}
\label{app:reporttype}

The results of the survey divided into overview report, subject report, timeline report, and vulnerability report are shown, respectively, in Table~\ref{tab:overview}, Table~\ref{tab:subject}, Table~\ref{tab:timeline}, and Table~\ref{tab:vulnerability}.
As we can see, the report type can heavily influence the utility score, while fluency and correctness are more consistent across the different templates.
In particular, with respect to Narrator, we can see a time reduction in report production of 27.8\% for overview reports, 20.7\% for subject reports, 6.2\% in timeline reports, and 46.9\% in vulnerability reports (on average 25.4\%).

\begin{table}[!htpb]
    \caption{Questionnaire results for the overview reports.}
    \begin{center}
        \begin{tabular}{|c|c|c|c|}
            \hline
            \textbf{Model}&\textbf{Fluency}&\textbf{Correctness}&\textbf{Utility}\\ \hline
            Narrator&2.92&2.77&109.2 min\\ \hline
            First Step AGIR&3.85&3.46&86.9 min\\ \hline
            \textbf{Final AGIR}&\textbf{4.08}&\textbf{4.23}&\textbf{78.9 min}\\ \hline
        \end{tabular}
    \label{tab:overview}
    \end{center}
\end{table}
\begin{table}[!htpb]
    \caption{Questionnaire results for the subject reports.}
    \begin{center}
        \begin{tabular}{|c|c|c|c|}
            \hline
            \textbf{Model}&\textbf{Fluency}&\textbf{Correctness}&\textbf{Utility}\\ \hline
            Narrator&2.85&2.92&91.2 min\\ \hline
            First Step AGIR&3.92&3.54&68.1 min\\ \hline
            \textbf{Final AGIR}&\textbf{4.00}&\textbf{4.15}&\textbf{72.3 min}\\ \hline
        \end{tabular}
    \label{tab:subject}
    \end{center}
\end{table}
\begin{table}[!htpb]
    \caption{Questionnaire results for the timeline reports.}
    \begin{center}
        \begin{tabular}{|c|c|c|c|}
            \hline
            \textbf{Model}&\textbf{Fluency}&\textbf{Correctness}&\textbf{Utility}\\ \hline
            Narrator&3.38&3.15&111.5 min\\ \hline
            First Step AGIR&3.46&3.38&114.6 min\\ \hline
            \textbf{Final AGIR}&\textbf{3.61}&\textbf{3.92}&\textbf{104.6 min}\\ \hline
        \end{tabular}
    \label{tab:timeline}
    \end{center}
\end{table}
\begin{table}[!htpb]
    \caption{Questionnaire results for the vulnerability reports.}
    \begin{center}
        \begin{tabular}{|c|c|c|c|}
            \hline
            \textbf{Model}&\textbf{Fluency}&\textbf{Correctness}&\textbf{Utility}\\ \hline
            Narrator&2.84&3.07&78.1 min\\ \hline
            First Step AGIR&3.84&3.53&48.8 min\\ \hline
            \textbf{Final AGIR}&\textbf{3.92}&\textbf{4.23}&\textbf{41.5 min}\\ \hline
        \end{tabular}
    \label{tab:vulnerability}
    \end{center}
\end{table}

\end{document}